\documentclass[aps,pre,twocolumn,groupedaddress,showpacs,floatfix]{revtex4}
\usepackage[american]{babel}
\usepackage[dvips]{graphicx}
\begin{document}
\title{Dependence of boundary lubrication on the misfit angle between the sliding surfaces}
\author{O.M. Braun}
\email[]{obraun.gm@gmail.com}
\affiliation{Institute of Physics, National Academy of Sciences of Ukraine,
46 Science Avenue, Kiev 252022 Ukraine}
\author{Nicola Manini}
\affiliation{Dipartimento di Fisica, Universit\`a degli Studi di Milano,
Via Celoria 16, 20133 Milano, Italy}
\date{June 10, 2010}
\begin{abstract}

Using molecular dynamics based on Langevin equations with a coordinate-
and velocity-dependent damping coefficient,
we study the frictional properties of a thin layer of
``soft'' lubricant (where the interaction within the lubricant is weaker
than the lubricant-substrate interaction) confined between two solids.
At low driving velocities the system
demonstrates stick-slip motion.
The lubricant may or may not be melted during sliding,
thus exhibiting either the ``liquid sliding'' (LS) or
the ``layer over layer sliding'' (LoLS) regimes.
The LoLS regime mainly
operates at low sliding velocities.
We investigate
the dependence of friction properties on the
misfit angle between the sliding surfaces
and calculate the
distribution of static frictional thresholds for a contact of polycrystalline
surfaces.
\end{abstract}
\maketitle

\section{Introduction}
\label{introduction}

The problem of boundary lubrication is very interesting from the physical
point of view and important for practical applications, but it is not
fully understood yet \cite{P0,BN2006}.
Conventional lubricants belong to the type of liquid (``soft'') lubricants,
where the amplitude of molecular interactions within the lubricant,
$V_{ll}$, is smaller than the lubricant-substrate interaction, $V_{sl}$.
Due to strong coupling with the substrates, lubricant monolayers cover the
surfaces, and protect them from wear.
A thin lubricant film, when its thickness is lower than about six molecular
layers, typically solidifies even if the conditions (temperature and
pressure) are those corresponding to the bulk liquid state.
As a result, the static friction force is nonzero, $f_s >0$, and the system
exhibits stick-slip motion, when the top substrate is driven through an
attached spring (which also may model the slider elasticity).
In detail, at the beginning of motion the spring elongates, the driving
force increases till it reaches the static threshold $f_s$.
Then a fast sliding event takes place, the spring relaxes, the surfaces
stick again, and the whole process repeats itself.
This stick-slip regime occurs at low driving velocities, while at high
velocities it turns into smooth sliding.

Since the pioneering work
by Thompson and Robbins \cite{TR1990,TR1990a},
who studied the lubricated system by molecular dynamics (MD),
the stick-slip is associated with the melting-freezing mechanism:
the lubricant film melts during slip and solidifies again at stick.
Such a sliding may be named the ``liquid sliding'' (LS) regime.
However, at low velocities the ``layer over layer sliding'' (LoLS) regime
sometimes occurs, where the lubricant keeps well ordered layered structure,
and the sliding occurs between these layers \cite{BN2006}.

In real systems the substrates are often made of the same material and may
even slide along the same crystallographic face, but can hardly be assumed
to be perfectly aligned, especially if the substrates have polycrystalline
structure.
In the majority of MD simulations, however, both substrates
are modelled identically, i.e., they have the same structure and
are perfectly aligned.
This fact may affect strongly the simulation results, as became clear after
predicting the so-called superlubricity, or structural
lubricity \cite{HS1990}.
For example, the ``dry'' contact (no lubricant) of two incommensurate rigid
infinite surfaces, produces null static friction, $f_s =0$
\cite{HS1990,MR2000,HMR1999,MWR2000b}.
If the surfaces are deformable, an analog of the Aubry
transition should occur with the change of stiffness of the substrates (or
the change of load \cite{L2002}): the surfaces are locked together for a
weak stiffness, and slide freely over each other for sufficiently high
stiffness (this effect was observed in simulation \cite{MR2000}).

In a real-world 3D contact, incommensurability can occur even for two
identical surfaces, if the 2D
surfaces are rotated with respect to each other.
Simulations \cite{HS1993,RS1996,Verhoeven04,Bonelli09,FasolinoInTrieste} do
show a large variation of friction with relative orientation of the two
bare substrates.
Similarly to the 1D Frenkel-Kontorova system, where the amplitude of the
Peierls--Nabarro barrier is a nonanalytic function of the
misfit parameter, in the 2D system the static frictional force
should be a nonanalytic function of the misfit angle between the two
substrates.
This was pointed out by Gyalog and Thomas in their study of the 2D
FK--Tomlinson model \cite{GT1997}.
However, surface irregularities as well as fluctuations of atomic positions
at nonzero temperature makes this dependence smooth and less pronounced.
For example, MD simulations \cite{QCCG2002} of the Ni(100)/Ni(100)
interface at $T=300$~K showed that for the case of perfectly smooth
surfaces, a $\pi/4$ rotation leads to a decrease in static friction by a
factor of $34 \div 330$.
However, if one of the surfaces is roughened with an amplitude 0.8~\AA,
this factor reduces to $4$ only, which is close to values observed experimentally.
M\"user and Robbins \cite{MR2000} noted that for a contact of atomically
smooth and chemically passivated surfaces, realistic values of the
stiffness usually exceed the Aubry threshold, 
thus one should 
expect $f_s =0$ for such a contact.
An approximately null static frictional force was indeed observed
experimentally in the contact of tungsten and silicon
crystals \cite{HSKM1997}.
More recently the friction-force microscopy experiment made by Dienwiebel
\textit{et al.}\ \cite{DVPFHZ2004} demonstrated a strong dependence of the
friction force on the rotation angle for a tungsten tip with an attached
graphene flake sliding over a graphite surface, where sliding occurs
between the graphene layers as relative rotation makes them incommensurate.

The case of lubricated friction was investigated by He and Robbins
\cite{HR2001s,HR2001k} for a very thin lubricant film (one monolayer or
less).
The dependences of the static \cite{HR2001s} and kinetic \cite{HR2001k}
friction on the rotation angle were calculated.
The authors considered the rigid substrates of fcc crystal with the (111)
surface and rotated the top substrate from $\phi=0$ to $\pi/6$.
It was found that static friction exhibits a peak at the commensurate angle
($\phi=0$) and then is approximately constant; the peak/plateau ratio is
about 7 (for the monolayer lubricant film, where the variation is the
strongest).
The kinetic friction varies slowly with a minimum at the commensurate angle
and a smooth maximum at $\phi \approx \pi/18 = 10^{\circ}$, changing by a
factor near two.
Also, the kinetic friction decreases with velocity at $\phi=0$, while it
increases at the other angles.

\textit{The goal of our work is a detailed MD study of stick-slip and
  smooth sliding for lubricated system with rotated surfaces.}
Compared to the work by He and Robbins \cite{HR2001s,HR2001k}, we study
thicker lubricant films, up to five atomic layers thick.
We explore a fairly basic model (see Sec.~\ref{model} and
Ref.~\cite{BP2001}. Interactions are of simple Lennard-Jones type,
typically each substrate consists of two layers with 12$\times$11 atoms,
arranged as a square lattice, and the lubricant has 80 atoms per layer)
which allows us a rather detailed study of the system dynamics for long
simulation times.
This model attempts to address the effects of relative crystal rotations in
generic lubricated sliding, without focusing on a specific system.
While the microscopic interactions are treated at a minimal level of
sophistication, we describe energy dissipation by means of a ``realistic''
damping scheme, with a damping coefficient in Langevin equation, which
mimics the energy exchange between the lubricant atoms and the substrate.
This is certainly important for smooth sliding, the kinetics of melting and
freezing processes and the ensuing metastable configurations which emerge
at stick during stick-slip regime.

Section~\ref{simulation} presents typical simulation results. The model
exhibits stick-slip at a low driving speed, which changes to smooth sliding
with increasing speed.
In the smooth sliding regime, as well as during slips in the stick-slip
regime, the system indeed exhibits either the LS or LoLS regime,
depending on the simulation parameters, and in particular the rotation
angle.
A new important result of the present study is that the LoLS regime should
be observed much more often than the LS regime.
Section~\ref{discussion} discusses and summarizes the obtained results.

\section{The model}
\label{model}

\begin{figure*}
\centerline{
\hfill(a)\includegraphics[clip,width=52mm,clip=]{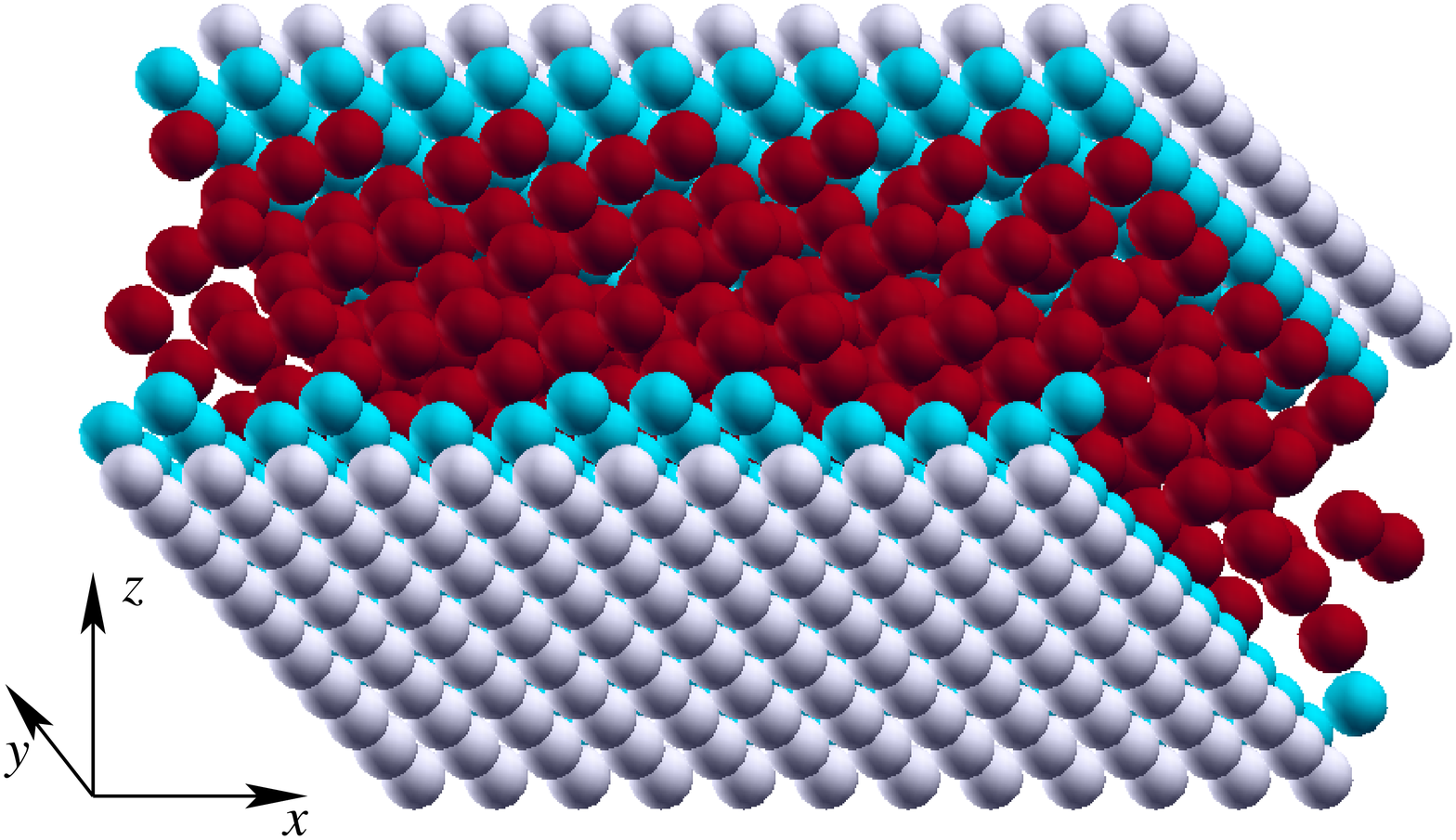}
\hfill(b)\includegraphics[clip,width=52mm,clip=]{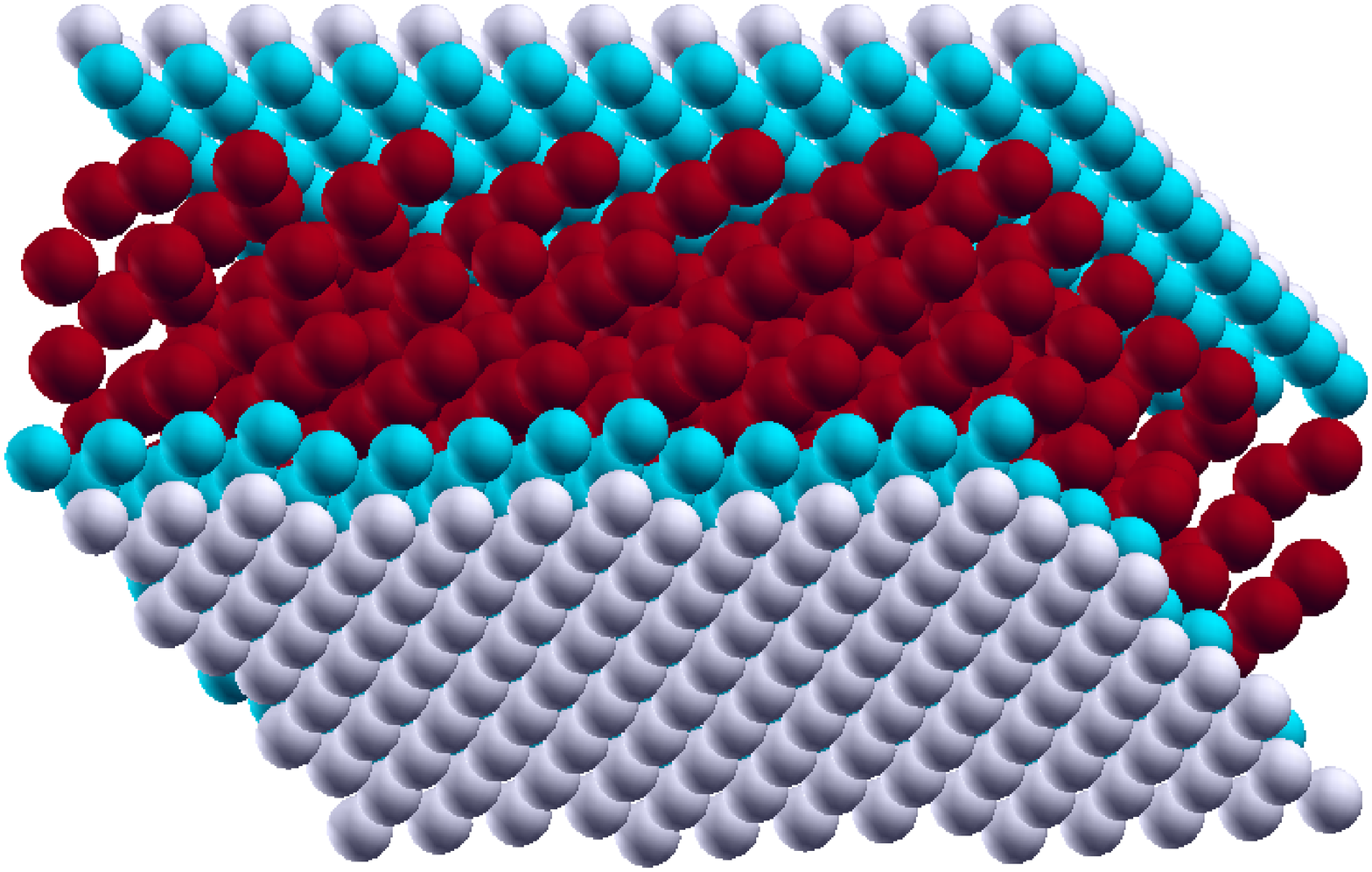}
\hfill(c)\includegraphics[clip,width=52mm,clip=]{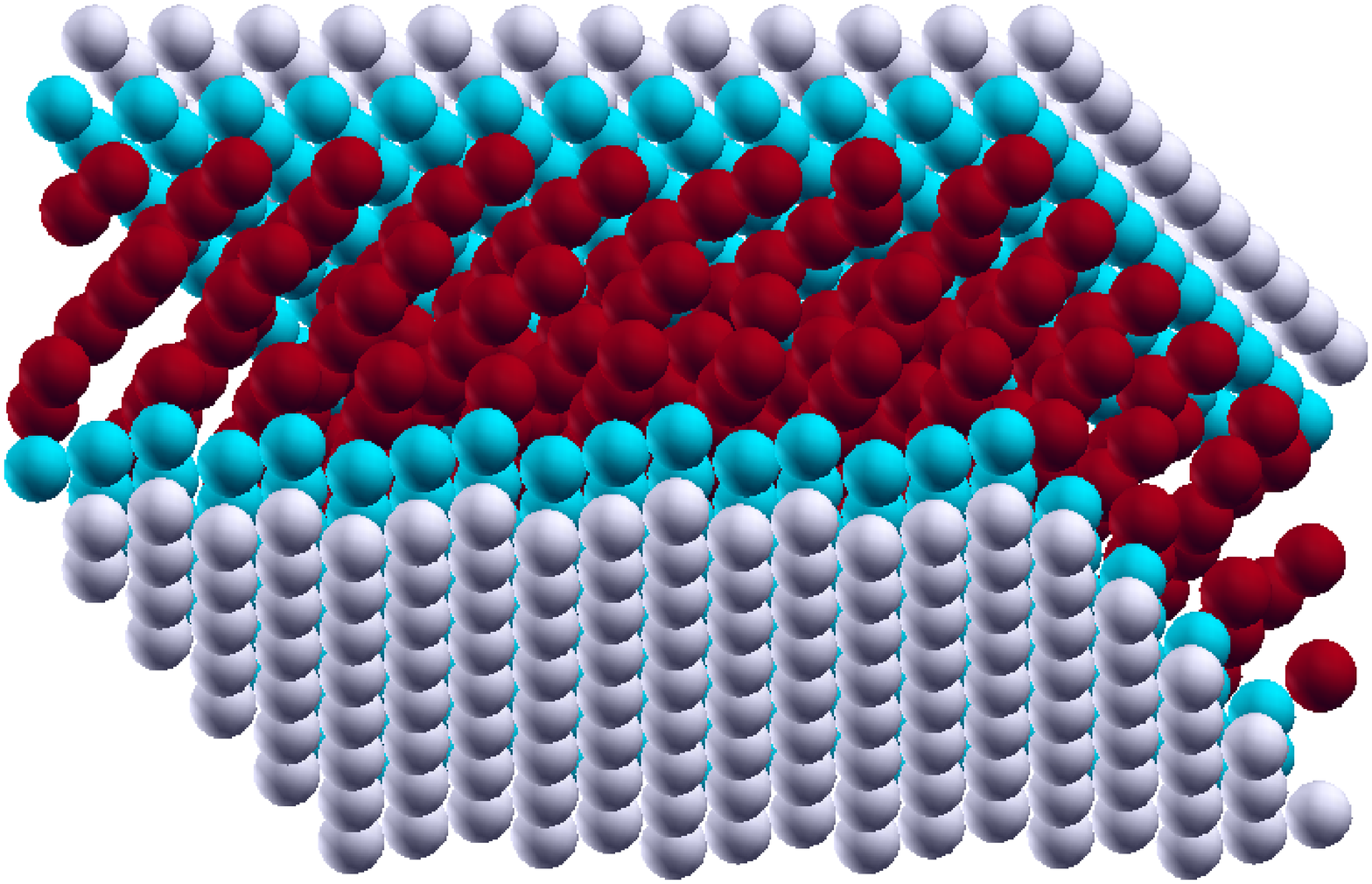}
\hfill}
\caption{\label{B06}(Color online)
Three typical stick configurations of the $N_l =3$ system during stick-slip
motion for misfit angles
(a)~$\phi=0^\circ$, (b)~$17.7^\circ$, and (c)~$28.7^\circ$.
Lubricant atoms are dark/red, deformable substrate atoms are clearer/blue,
and fixed substrate atoms are white.
Driving speed is $v=0.03$, load $f_l=0.1$.
}
\end{figure*}

As we address rather general properties of lubricated friction,
we explore a relatively simple model in simulation.
This allows us to span a wider range of sliding velocities and longer
simulation times as well as to analyze the atomic trajectories in greater
detail than by simulating fully realistic force fields appropriate to a
specific system.
The MD model was introduced and described in previous work
\cite{BN2006,BP2001}; therefore, we only discuss here its main features
briefly.
We study a film composed by few atomic layers and confined between two
substrates, ``bottom'' and ``top''.
As illustrated in Fig.~\ref{B06}, each substrate consists of two layers.
The external one is rigid, while the dynamics of the atoms belonging to the
inner layer, the one directly in contact with the lubricant, is fully
included in the model.
The rigid part of the bottom substrate is held fixed, while the top
substrate is mobile in the three space directions $x,y,z$.

All atoms interact with pairwise Lennard-Jones potentials
\begin{equation}
V(r)=V_{\alpha \alpha^{\prime}} \left[ \left( {r_{\alpha \alpha^{\prime}}}/{r}
\right)^{12} -2 \left( {r_{\alpha \alpha^{\prime}}}/{r} \right)^{6} \right],
\label{m0}
\end{equation}
where $\alpha, \alpha^{\prime} = s$ or $l$ for the substrate or lubricant
atoms respectively, so that the interaction parameters
$V_{\alpha \alpha^{\prime}}$ and $r_{\alpha \alpha^{\prime}}$
depend on the type of atoms.
Between two substrate atoms we use $V_{ss} = 3$ and an equilibrium distance
$r_{ss} = 3$.
The interaction between the substrate and the lubricant is much weaker,
$V_{sl} = 1/3$.
For the ``soft'' lubricant itself, we consider $V_{ll} = 1/9$ and an
equilibrium spacing $r_{ll} = 4.14$, which is poorly commensurate with
$r_{ss}$.
The equilibrium distance between the substrate and the lubricant is
$r_{sl} = \frac{1}{2}(r_{ss} + r_{ll})$.
For the long-range tail of all potentials we adopt a standard truncation to
$r \le r^* = 1.49 \, r_{ll}$.
The atomic masses are $m_l = m_s = 1$.
The two substrates are pressed together by a loading force $f_l$ per
substrate atom (typically we used the value $f_l = 0.1$).
%
All parameters are given in dimensionless units defined in
Ref.~\cite{BP2001}, for example the model units for force is $V_{ss} /
r_{ss}$.
%
The chosen parameters correspond roughly to a typical system where energy
is measured in electronvolts and distances in \aa ngstr\"oms, so that
forces are in the nanonewton range.

The main difference between our technique and other simulations of confined
systems
lies in the dissipative coupling with the heat bath,
representing the bulk of the substrates.
%
We use Langevin dynamics with a position- and velocity-dependent damping
coefficient $\eta (d_i,v_i)$, which is designed to mimic realistic
dissipation, as discussed in Refs.~\cite{BN2006,BP2001}.
In a driven system the energy pumped into the system must be removed from
it.
In reality energy losses occurs through the excitation of degrees of
freedom not included in the calculation, namely energy transfer into the
bulk of the substrates.
To model this fact, damping should occur primarily when a moving atom comes
at a small distance $d_i$ from either substrate.
Moreover, the efficiency of the energy transfer depends on the velocity
$v_i$ of the atom because $v_i$ affects the frequencies of the motions that
it excites within the substrates.
The damping is written as $\eta(d,v)= \eta_1(d) \, \eta_2(v)$ with
$\eta_1(d) = 1 - \tanh[ (d - d^*)/d^*] $, where $d^*$ is a characteristic
distance of the order of one lattice spacing.
The expression of $\eta_2(v)$ is deduced from the results known for the
damping of vibrations of an atom adsorbed on a crystal surface (see
Ref.~\cite{BF2002} and references therein).

In simulations we explore the ``spring'' algorithm, where a spring is
attached to the rigid top layer, and the spring end is driven at a constant
velocity.
%
%
We apply periodic boundary conditions (PBC) in the $x$ and $y$ directions.
The geometric construction of the rotated substrate is explained in the
Appendix~\ref{appendix}.
In simulation, it is simpler to rotate the bottom substrate only.
The initial configuration of the lubricant is prepared as a set of
$N_l$ ($N_l =1$, 2, 3, and 5) closely packed atomic layers.
Most simulations include $80$ atoms in each lubricant layer, although we
increased the system size by up to 16 times to check for size effects.
%
The system is then annealed, i.e., the temperature is raised adiabatically
to $T \sim 0.6$, which exceeds the melting temperature $T_m$ ($T_m \simeq
0.1$ for our lubricant \cite{BP2003}) and then decreased back to the
desired value.
After preparation of the annealed configuration, we perform a standard
protocol of runs: starting from the high-speed $v=1$ LS regime,
corresponding to a sliding speed comparable to the sound velocity of the
lubricant in its solid state, we reduce the driving velocity $v$ in steps
down to the value $v=0.01$ (which produces stick-slip motion for most
employed simulation parameters), and then increase $v$, up to $v=3$.
Typical MD snapshots are shown in Fig.~\ref{B06}.

To estimate $f_s$ in the stick-slip regime we select the $v=0.03$ runs and
take an average of the peak spring force immediately prior to slip.
A lower sliding speed would lead to slightly larger friction, due to longer
aging of pinning contacts, but would require longer simulation times to
record the same number of stick-slip events.
The moderate speed $v=0.03$ realizes a fair compromise, which in practical
simulation times produces a $\sim 10\%$ underestimate of $f_s$ with respect
to its value at adiabatically slow sliding.
To find the kinetic friction force $f_k$ in the smooth sliding regime,
we average the spring force over the whole run.

\section{Simulation results}\label{simulation}

\begin{figure}
\includegraphics[width=8cm,clip=]{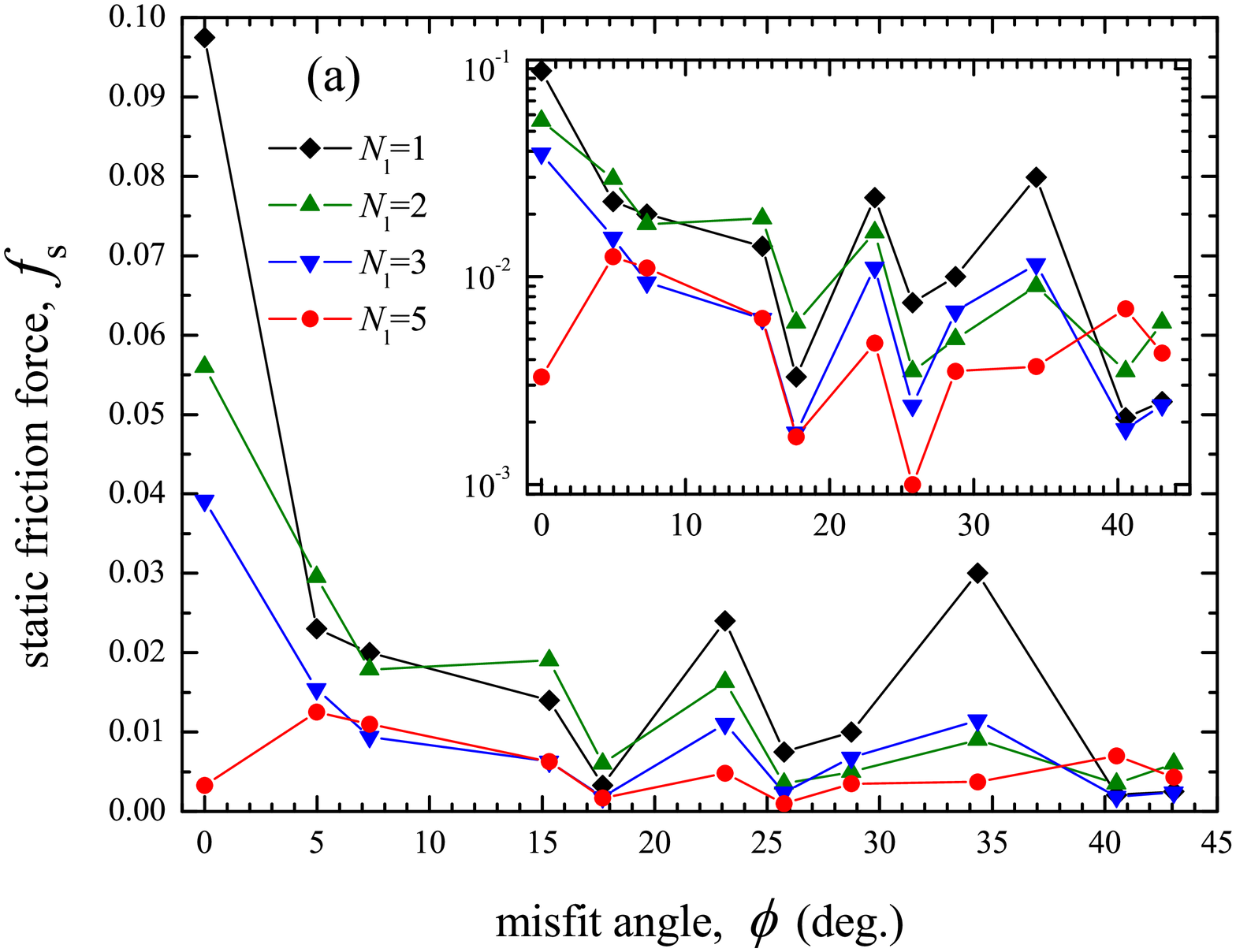}
\includegraphics[width=8cm,clip=]{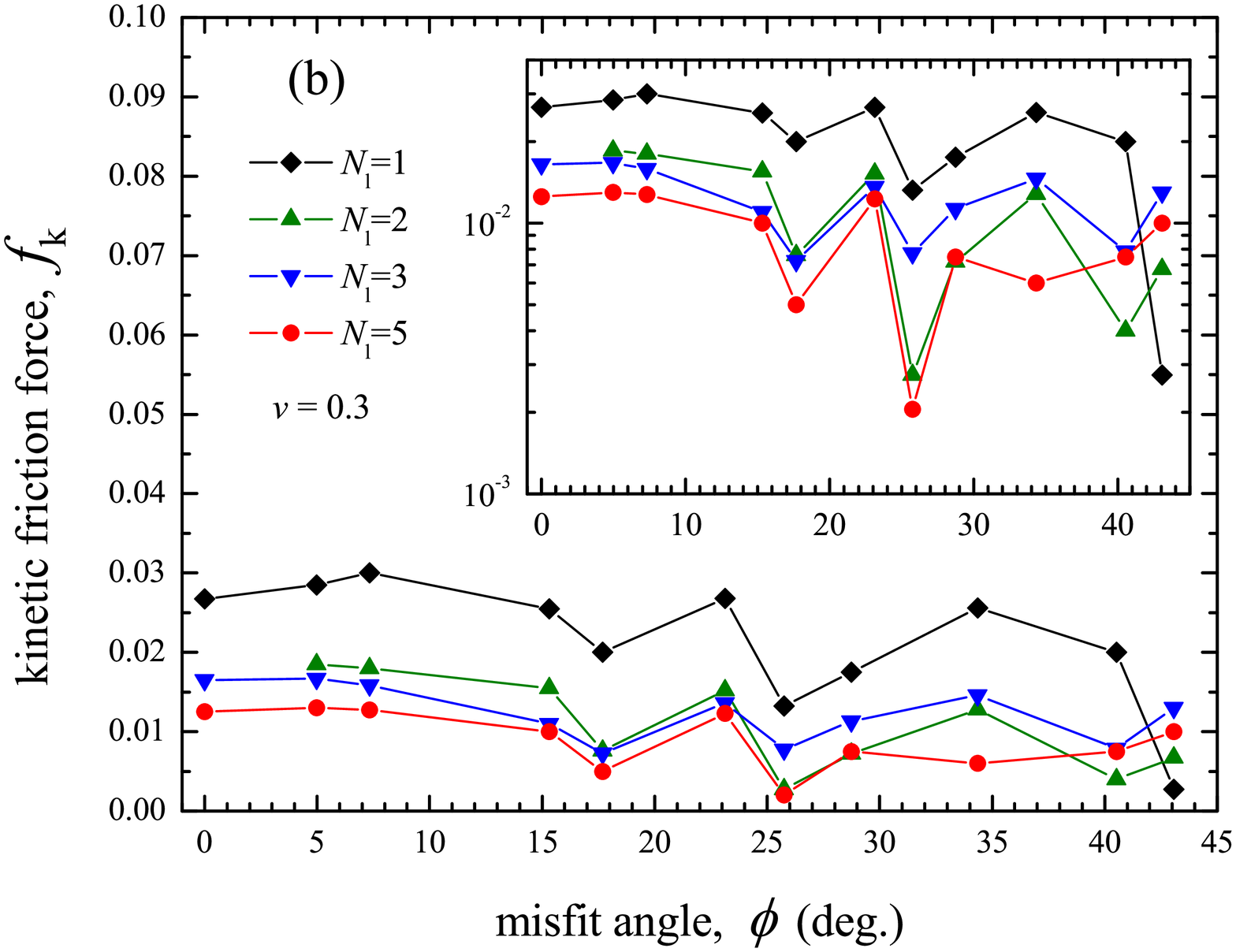}
\caption{\label{C17all}(Color online)
Static (panel a, stick slip at $v=0.03$) and kinetic (panel b,
smooth sliding with $v=0.3$) friction forces for several values of the
misfit angle $\phi$ for $T=0$ and different thicknesses of the lubricant
film: $N_l =1$ (diamonds), 2 (green up triangles), 3 (blue down triangles), and~5
(red circles).
Insets display the same dependences in logarithmic scale.
All forces here and in the following figures are reported in the natural
model units, i.e.\ in units of $V_{ss} / r_{ss}$.
}
\end{figure}

The simulation results are summarized in Fig.~\ref{C17all}.
Qualitatively, the results agree with those
of He and Robbins \cite{HR2001s,HR2001k},
with stick-slip motion at low driving velocities and  smooth sliding for
$v \gtrsim 0.1$.
%
%
%
At zero substrate temperature, the static friction can vary with the misfit
angle $\phi$ by two orders of magnitude, and the kinetic friction for
smooth sliding at low velocity (e.g. as shown for $v=0.3$ in
Fig.~\ref{C17all}b) by more than one order of magnitude.
For the thin lubricant films, $N_l \le 3$, the static friction peaks
for perfectly aligned substrates, $\phi =0$.
This does not occur any more for thicker films.
%
The friction achieves sharp minima at the 
angles
$\phi=17.7^{\circ}$, $25.7^{\circ}$ and $40.5^{\circ}$
as will be discussed below.
%
At large driving velocities, $v \ge 1$ (which is in fact huge,
comparable to the solid-lubricant sound velocity) the lubricant film is
completely molten, and friction becomes almost independent of $\phi$.

\begin{figure}
\includegraphics[width=8cm,clip=]{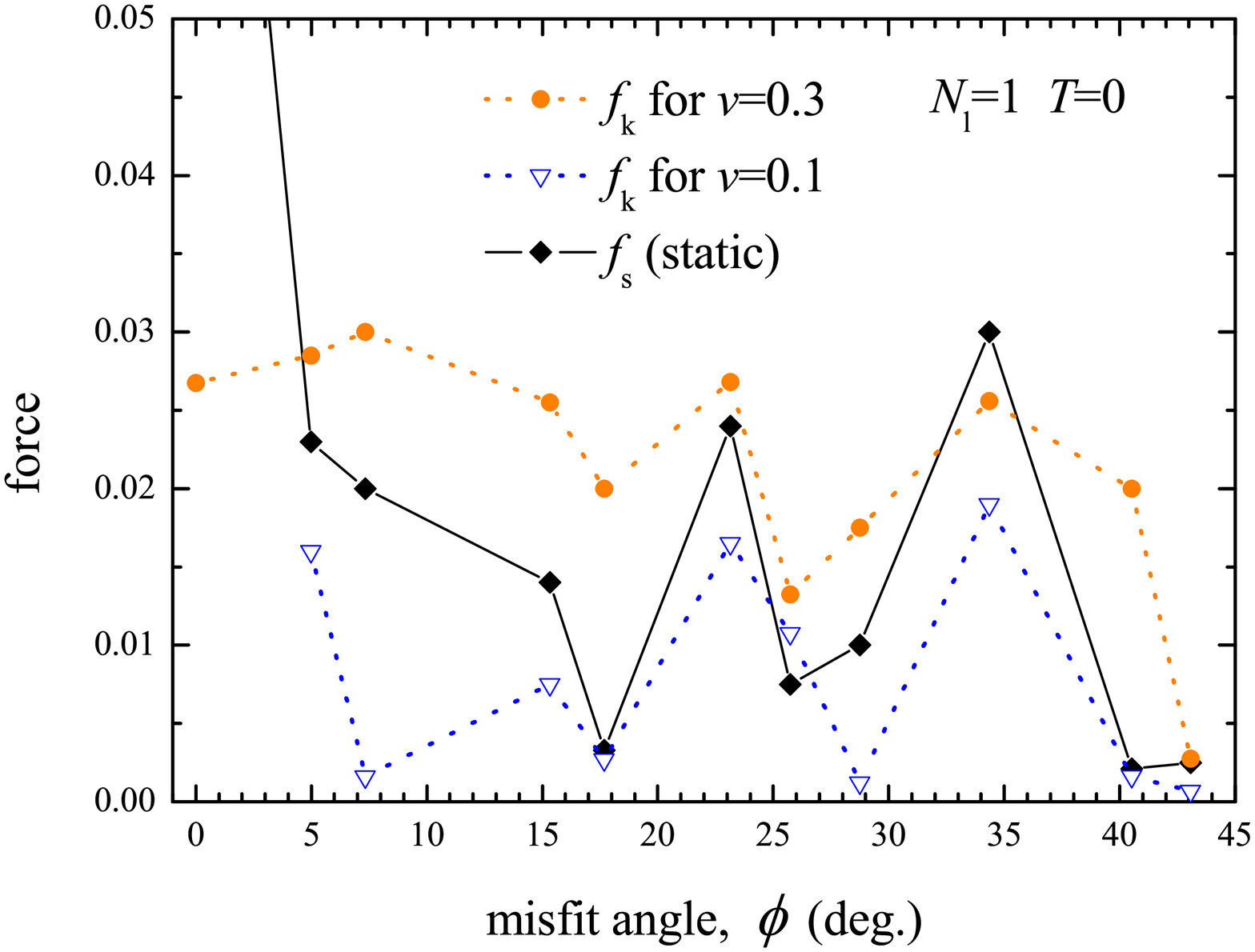}
\caption{\label{C17c}(Color online)
Friction force as a function of the misfit angle $\phi$
for the one-layer lubricant film 
at $T=0$.
Diamonds correspond to static friction,
triangles to kinetic friction at $v=0.1$, and
circles to kinetic friction at $v=0.3$.
}
\end{figure}

The variation of friction with $\phi$ is the most pronounced for the
one-layer lubricant film.
The simulation results for this thinnest film are presented in
Fig.~\ref{C17c}. 
The static and low-speed ($v \le 0.3$) kinetic friction force displays
sharp minima for the angles $\phi=17.7^{\circ}$, $25.7^{\circ}$ and
$40.5^{\circ}$.
For these ``special'' angles, the lubricant film remains ordered and slides
together either with the top or the bottom substrate both during slips and
in smooth sliding (we call this regime as the ``solid sliding'', or SS).
Of course the motion is not rigid but corresponds to a ``solitonic''
sliding mechanism
\cite{BraunBook,Vanossi06,Vanossi07PRL}.
For the other angles studied, at stick configurations the film orders
locally, with a structure adjusted partly to the bottom and partly to
the top substrates, while during slips, as well as at smooth sliding, the
film is 2D-melted (LS regime).

\begin{figure}
\includegraphics[width=8cm,clip=]{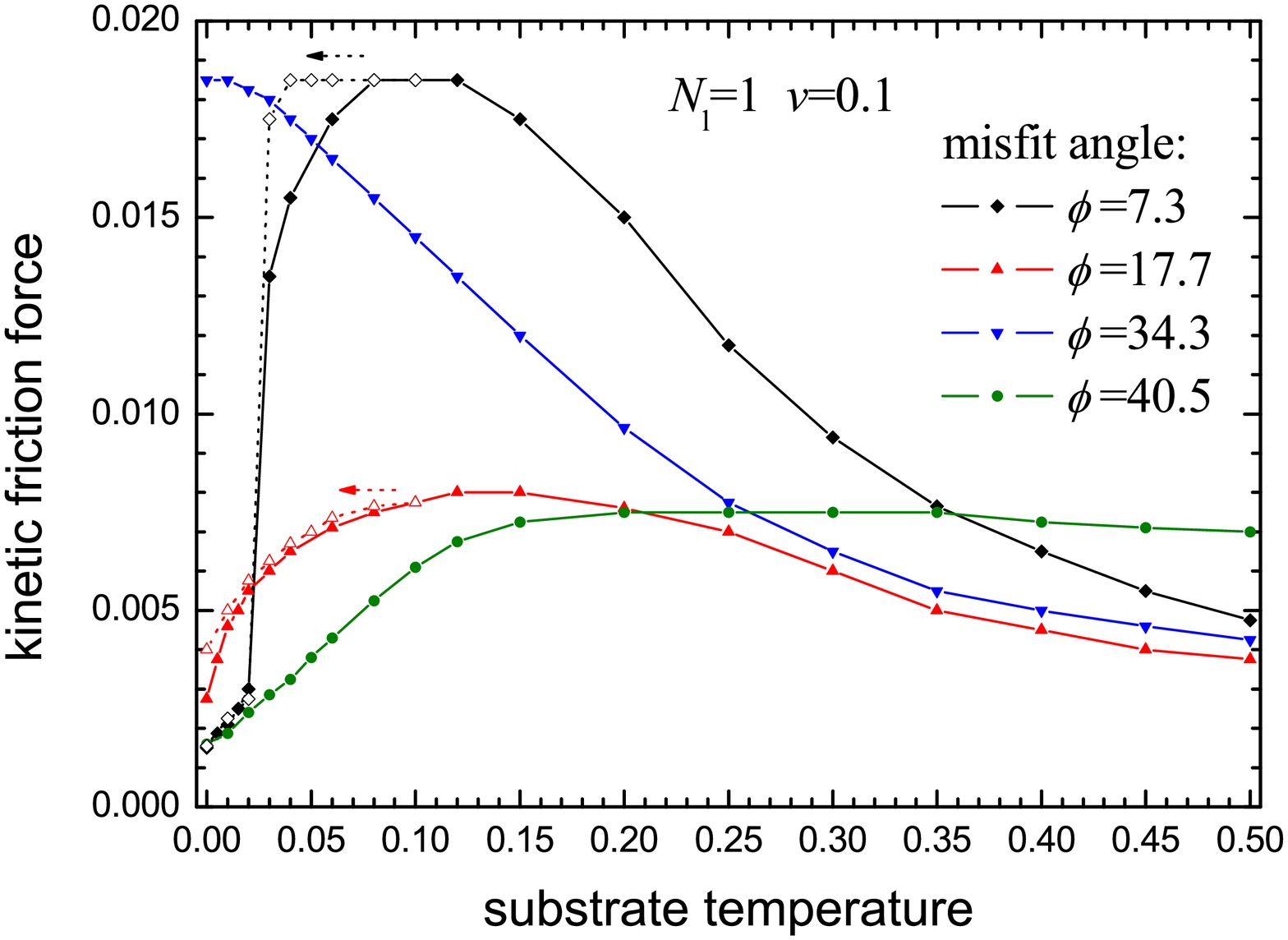}
\caption{\label{fw1kT}(Color online)
Kinetic friction force as a function of temperature
(in natural units of $V_{ss}/(3 k_{\rm B})$, where $k_{\rm B}$ is the
Boltzmann constant)
computed at $v=0.1$ for the misfit angles
$\phi=17.7^{\circ}$ and $\phi=40.5^{\circ}$ (solitonic),
$\phi=34.3^{\circ}$ (high friction), and
$\phi=7.3^{\circ}$ (fragile solitonic)
for the $N_l =1$ system.
}
\end{figure}

The dependence of the kinetic friction force on temperature is shown in
Fig.~\ref{fw1kT}.
For all ``non-special'' angles, when the static friction is relatively
high, the kinetic friction decreases with temperature (e.g., see the
dependence for $\phi=34.3^{\circ}$ in Fig.~\ref{fw1kT}), reflecting the
standard thermolubric effect
\cite{P0,BN2006,Gnecco00,Sang01,Riedo03,Gnecco03,Sills03,Jinesh08,Krylov08}
due to temperature-assisted barrier overcoming.
However, for all special angles producing the SS regime, the behavior is
different: thermal fluctuations perturb a rather delicate solitonic motion,
leading to an initial increase of friction with $T$.
Such a behavior occurs also for thicker films.
For the misfit angle $\phi\simeq 7.3^{\circ}$ we observe that at $v=0.1$ the
one-layer film reaches an exceptional sliding state characterized by very
low friction due to SS; this type of superlubricity is not typical, and it
is quickly destroyed with the increase of either temperature
(Fig.~\ref{fw1kT}) or velocity (Fig.~\ref{C17c}).
Note also that the superlubric SS regime is recovered significantly after
it is abandoned as temperature is lowered (dotted line, open symbols), thus
opening a nontrivial hysteretic loop in the thermal cycle.

\begin{figure}
\includegraphics[width=8cm,clip=]{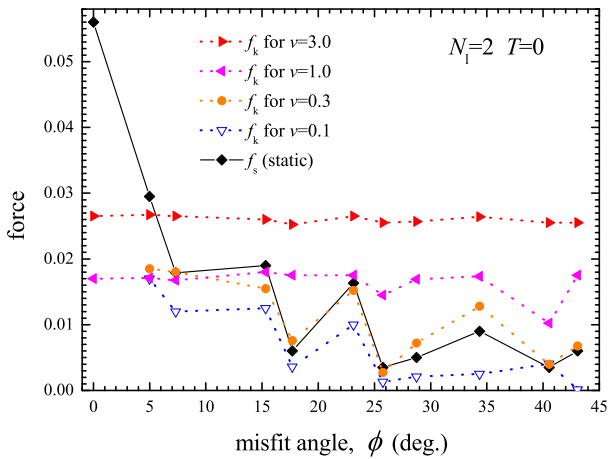}
\caption{\label{C17b}(Color online)
Friction forces versus the misfit angle $\phi$ for the $N_l =2$ system.
Diamonds and solid curve describe the static friction, dotted curves show
the kinetic friction force at different driving speeds, up to values
comparable to the lubricant sound velocity.
}
\end{figure}

Consider now a thicker lubricant film with $N_l=2$ (see Fig.~\ref{C17b}).
For $v \geq 1$ we observe the LS regime, where the lubricant is 3D molten
(except the ``special'' angles $\phi=25.7^{\circ}$ and $40.5^{\circ}$ at $v=1$,
where we have LoLS between the two attached lubricant layers).
At lower velocities, $v \leq 0.3$, the behavior is as follows.
For $\phi =0$ at stick, the two ordered lubricant layers are ordered and
attached to the corresponding substrates, but are 3D melted at slips.
For all other angles, the LoLS regime operates during slips:
for $\phi < 25^{\circ}$ the attached layers are 2D molten, while
for $\phi > 28^{\circ}$ the attached layers remain ordered.
Finally, for the angles $\phi=17.7^{\circ}$, $25.7^{\circ}$ and
$40.5^{\circ}$ the friction forces exhibit deep minima produced by the SS
solitonic mechanism.

\begin{figure}
\includegraphics[width=8cm,clip=]{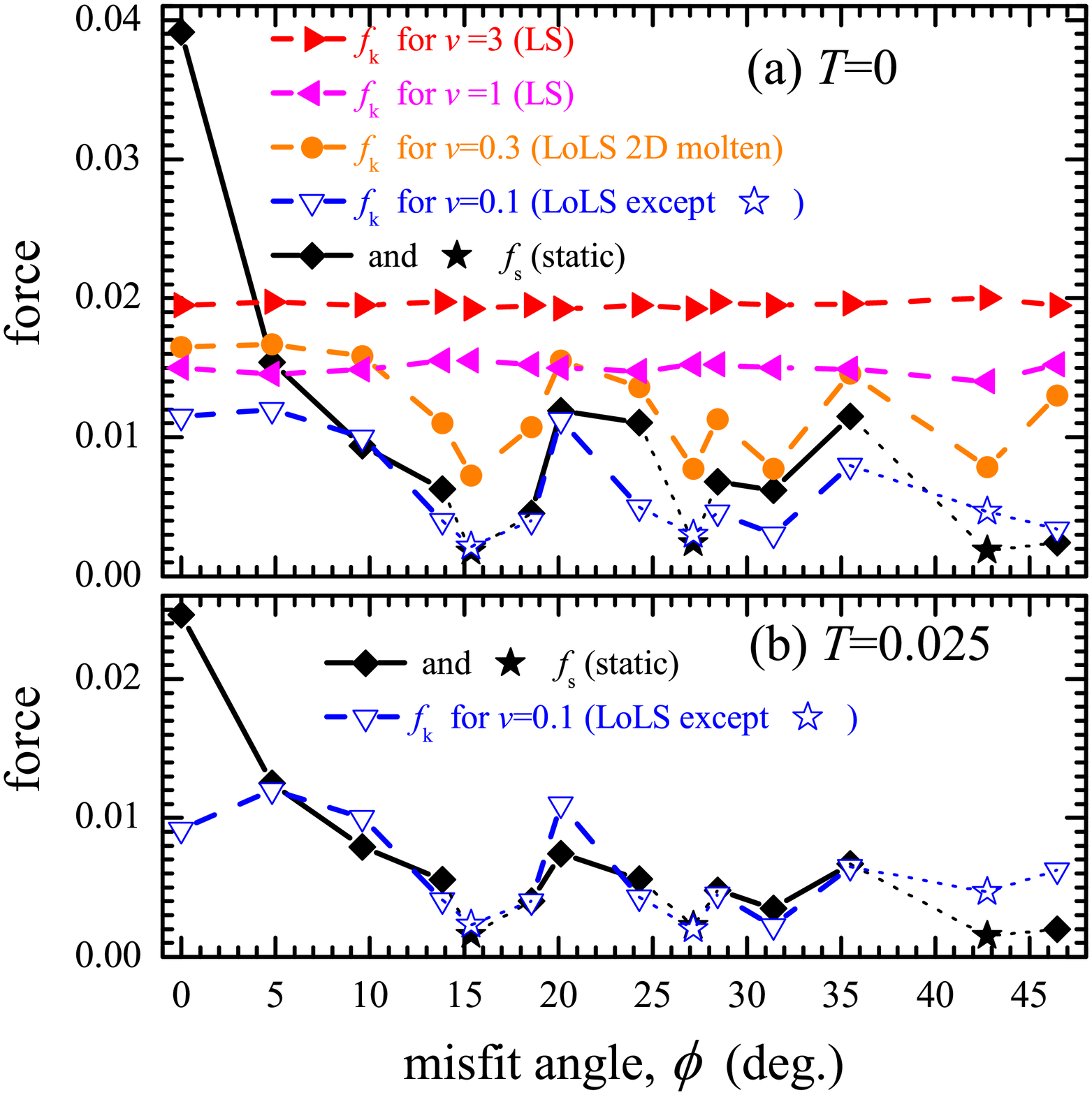}
\caption{\label{C15}(Color online)
Friction forces as functions of the misfit angle $\phi$ for the three-layer
system.
Diamonds and solid line represents static friction;
stars and dotted lines mark the regime of solitonic ``solid sliding'' with very low
friction (``superlubricity'').
Dashed lines show the
kinetic friction force at different driving velocities: $v=0.1$ (blue down
triangles), 0.3 (orange circles), 1 (cyan left triangles) and 3 (red right triangles).
Panel (a) is for $T=0$, and (b) is for the ``room'' temperature $T=0.025$.
}
\end{figure}

We come now to describe the results for the $N_l =3$ system as a
prototypical lubricant of mesoscale thickness.
The static friction, as well as the kinetic friction in the LoLS regime,
can change by more than one order in magnitude when the misfit angle
varies, as illustrated in Fig.~\ref{C15}.
$\phi=0$ produces the highest friction like for thinner lubricants.
The smooth sliding, as well as slips during stick-slip, correspond to
either the LoLS or the LS regime.
For all other angles $\phi \ne 0$, {\em sliding always corresponds to the
  LoLS regime} at low driving velocities $v \leq 0.3$.
Contrary to the $\phi=0$ case, now sliding is typically asymmetric and
takes place at one interface only, between the middle layer and one of the
attached layers, so that the middle lubricant layer sticks with either the
top or the bottom substrate.
The middle layer may remain ordered during sliding or, for some values of
$\phi$, it is 2D-melted; in the latter case, the friction force is higher.
For the special misfit angles $\phi=17.7^{\circ}$, $25.7^{\circ}$ and
$40.5^{\circ}$ identified by stars in Fig.~\ref{C15}, we again observe the
``superlubricity'' characterized by the very low friction.
In these cases, the lubricant film remains solid and ordered during
sliding, and moves as a whole with the top or bottom substrate, in a
SS sliding.
However, the lubricant is not rigid during motion, thus enhancing the
``solitonic'' mechanism.

The results described above, remain qualitatively the same at nonzero
temperatures $T<T_m$.
For example, Fig.~\ref{C15}b shows the dependence of friction force on
$\phi$ for the ``room'' temperature $T=0.025$.
Both the static and kinetic (for the LoLS regime) friction forces decrease
when $T$ increases.
However, for the SS regime, the behavior is different -- fluctuations due
to temperature perturb the 
solitonic motion, leading to an increase of friction.

\begin{figure}
\includegraphics[width=8cm,clip=]{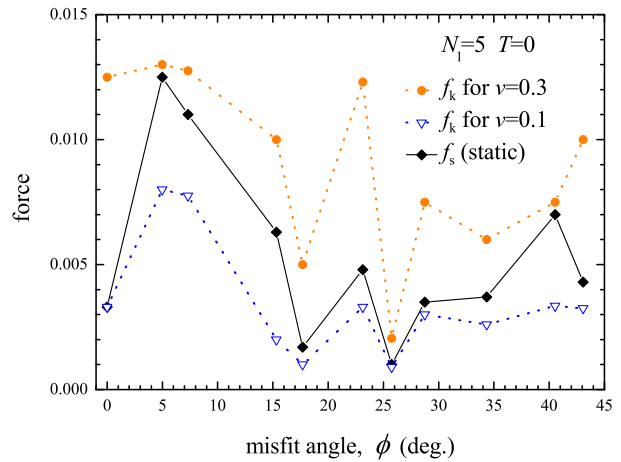}
\caption{\label{C17a}(Color online)
Friction forces versus the misfit angle $\phi$ for the $N_l =5$ system.
Diamonds and solid curve describe the static friction, dotted curves show
the kinetic friction force at driving velocities $v=0.1$ (triangles) and
0.3 (circles).
}
\end{figure}
The thicker lubricant film, $N_l =5$, behaves similarly to $N_l =3$, as
illustrated in Fig.~\ref{C17a}.
The main difference is the lack of a maximum in $f_s (\phi)$ at $\phi=0$.
Again, for the misfit angles $\phi=17.7^{\circ}$ and $25.7^{\circ}$ we
observe ``superlubric'' sliding.
%
For smooth sliding with $v \leq 0.3$ as well for slips during stick-slip
motion, we observe either the LoLS regime, where the three central layers
are 2D melted and sliding occurs between the middle layers (e.g., between
layers 1-2 or 2-3), or the LS regime, where all three middle layers are
3D-molten.
The LoLS regime marks the dips in $f_k$, while LS generates larger $f_k$,
as occurs in the angular intervals $0^\circ \le \phi \le 10^\circ$ and
$20^\circ \le \phi \le 25^\circ$.
For larger velocities, $v \geq 1$, all five lubricant layers are melted
and the LS regime operates in full.


\begin{figure}
\includegraphics[width=8cm,clip=]{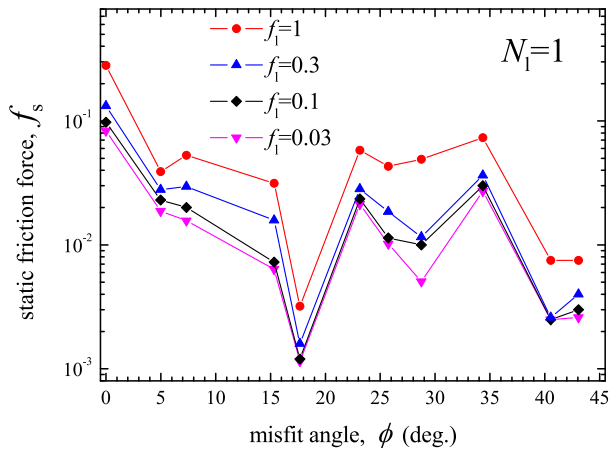}
\caption{\label{load1} (Color online)
Friction force versus the
misfit angle $\phi$ for the $N_l =1$ system, driven at $v=0.01$,
for different values of the applied load $f_l$.
}
\end{figure}

Figure~\ref{load1} reports the friction force for 4 values of the applied
load.
These calculations demonstrate that the dependency of friction on the
substrate rotation is very similar for different loads, with a general
increase of friction with load.
We obtain very similar results for $N_l=2$ and 3.
However, for thicker films, $N_l=5$, the situation changes dramatically at
high load ($f_l=1$): the film rearranges into a
closely packed four-layer
configuration which remains solid under sliding.
As a consequence, the peak structure of friction as a function of $\phi$
changes as well, because the lubricant structure acquires more atoms per
layer and changes symmetry. 

\begin{figure}
\includegraphics[width=8cm,clip=]{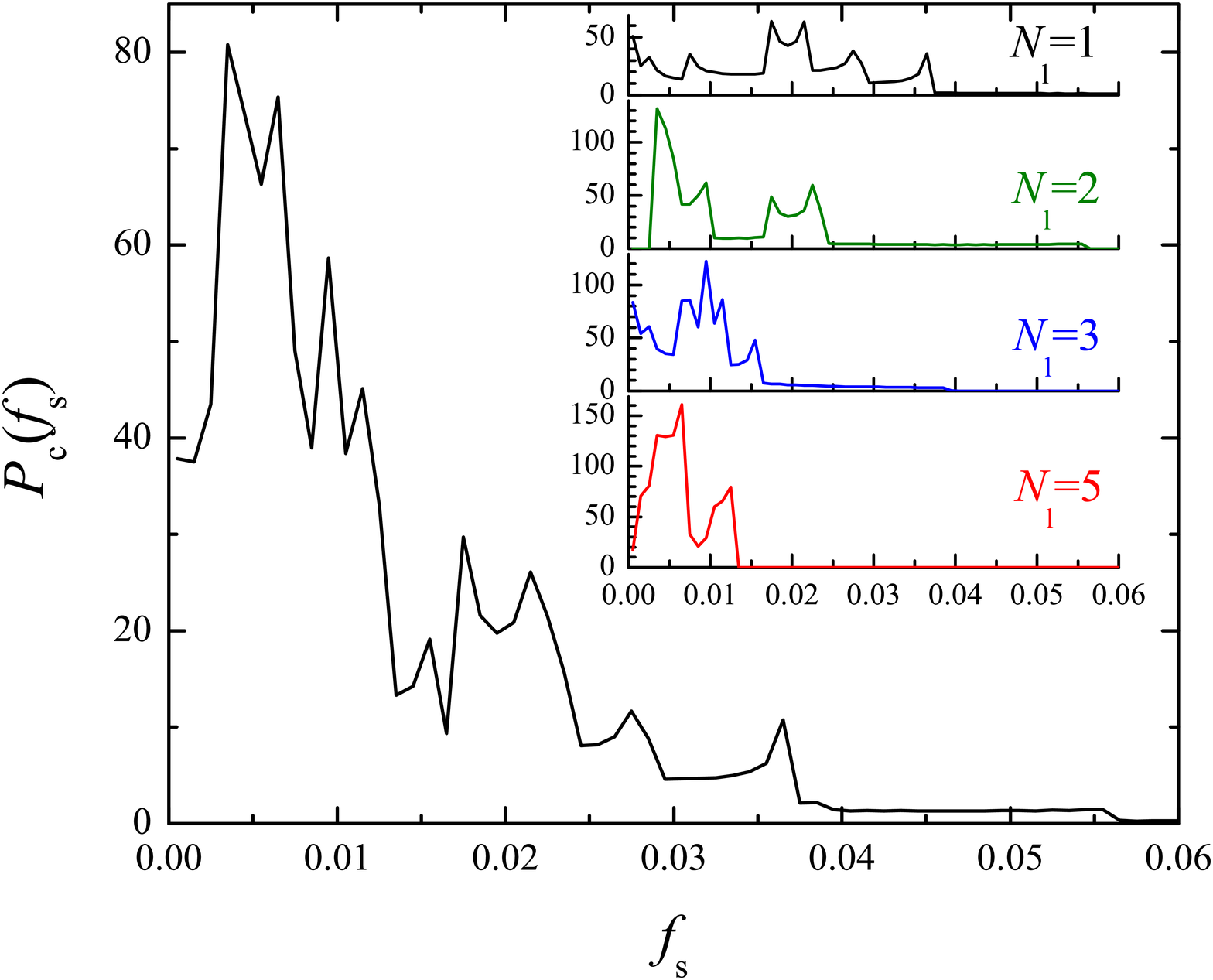}
\caption{\label{static1235} (Color online)
The probability distribution of static thresholds for $N_l =1$, 2, 3 and~5
(insets) and the averaged distribution $P_c (f_s)$.
These distributions are obtained in the assumption that all misfit angles
$\phi$ are equally likely.
The calculated function $P_c (f_s)$ may be approximately fitted
by the distribution $P_c (f_s) = \bar{f}^{-2} f_s \exp ( -f_s
/\bar{f})$ with $\bar{f} =0.005$.
}
\end{figure}

Because the static friction varies over such a large interval, its specific
value for a given misfit angle $\phi$ has little importance.
Indeed one could hardly control the misfit angle in a real system, except
in especially favorable situations \cite{DVPFHZ2004}.
Moreover, a system where the sliding surfaces have an ideal crystalline
structure oriented with a controllable $\phi$ is exceptional.
%
Real surfaces usually have areas (domains) with different orientation.
For polycrystalline substrates, it is reasonable to assume that all angles
are presented with equal likeliness.
It may then be more interesting to examine the probability distribution of
$f_s$ values, regardless of $\phi$.
The insets of Fig.~\ref{static1235} report the histograms of forces as
resulting from our simulations of different thicknesses.

Besides, if we also assume that the lubricant film is not uniform but has
different thickness at different places (it is certainly so if the
surfaces have some roughness), then we can average over different
thicknesses; the resulting distribution $P_c(f_s)$ is shown
Fig.~\ref{static1235}.
It is precisely this distribution which represents the main output of our
MD simulations, as it then allows us to predict tribological behavior of
the system with the help of a master-equation approach \cite{BP2008}.
%
%
The calculation summarized in Fig.~\ref{static1235} are carried out at
$f_l=0.1$.  The effect of a load increase would be mainly to scale the
$P_c(f_s)$ distribution, so that it would peak at larger friction.
The resulting distribution displays several spikes which are likely due to
the fixed size of the simulated contact.
We will investigate the role of contact size on the $P_c(f_s)$ distribution
using a simplified model, in a separate publication \cite{friction_rigid}.

\section{Discussion and Conclusions}
\label{discussion}

The main results of the present work can be summarized as follows:
(\textit{i}) The relative rotation of sliding contacts in a lubricated context
promotes LoLS more frequently then the standard LS.
(\textit{ii}) For a few special angles LoLS leads to superlubric sliding
completely analogous of the unlubricated sliding of misaligned perfect
crystalline substrates.  This superlubric regime is however delicate and
can be suppressed by small relative rotations of the substrates, by
temperature, or by velocity-induced local heating.
(\textit{iii}) In a regime of boundary lubrication, friction forces do vary
quite significantly with the relative substrate relative orientation
$\phi$, even when the lubricant film becomes several atomic layers thick.
(\textit{iv}) To describe macroscopic friction in a context of
multiasperity contact, where relative orientation is not really under
control, the most important information to be extracted from MD simulations
is a probability distribution $P_c(f_s)$ rather than specific values of the
static friction force $f_s$.

The present calculation is consistent with a rapidly (approximately
exponentially) decaying distribution $P_c(f_s)$, up to a cutoff force
related to the average contact size.
When temperature can promote rotations, beside the standard reduction of
friction at low speed due to thermal crossing of barriers, thermal
fluctuations could affect the barrier distribution itself by suppressing
small barriers in favor of higher ones \cite{FDFKU2008}.

\acknowledgments

We wish to express our gratitude to B.N.J.\ Persson for helpful discussions.
This research was supported in part by a grant from the Cariplo Foundation
managed by the Landau Network -- Centro Volta,
and by the
Italian National Research Council (CNR, contract ESF/EUROCORES/FANAS/AFRI),
whose contributions are gratefully acknowledged.

\appendix*
\section{The construction of the rotated substrate}
\label{appendix}

\begin{figure}
\includegraphics[width=8cm,clip=]{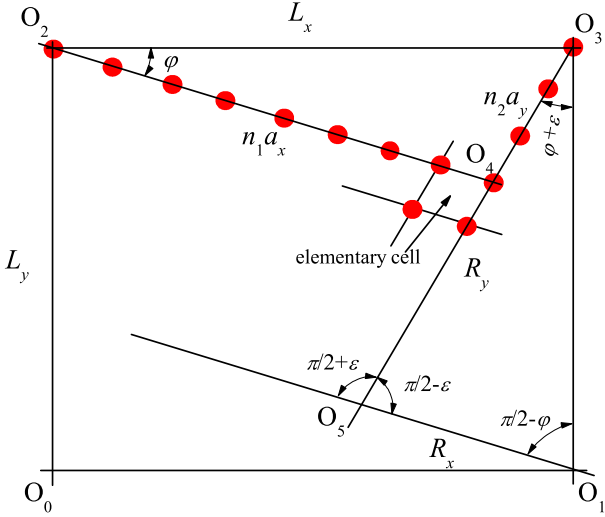}
\caption{\label{C16}(Color online)
Construction of the rotated lattice.
}
\end{figure}
In this Appendix we report the construction of the substrate rotated by a
given misfit angle $\phi$.
The main problem here is to obtain numerous misfit angles $\phi$ which
satisfy PBC in $x$ and $y$ directions simultaneously, while maintaining a
constant simulation size, and rectangular PBC
(for the square shape of the simulation cell
the construction is much simpler, e.g., see Ref.~\cite{GT1997}).
The idea of the construction is demonstrated in Fig.~\ref{C16}.
The substrate is arranged according to a square lattice with lattice
constant $a_s$, and the simulation cell area is $L_x \times L_y = M_x a_s
\times M_y a_s$.
The rotated bottom lattice is constructed as a set of parallelograms, so
that the elementary cell of the new lattice has size $a_x \times a_y$ with
a base angle $\pi/2 + \varepsilon$.
In the perfect case we would have $a_x = a_y = a_s$ and $\varepsilon =0$.
However, to satisfy the PBC, the rotated lattice has to be distorted from the
ideal square shape, and the idea is to reduce this distortion to a minimum.

The rotated lattice is defined by two integers $n_1$ and $n_2$ (see
Fig.~\ref{C16}) which determine the rotation angle $\varphi$.
For example, the choice $n_1=M_x$ and $n_2=0$ or $n_1=0$ and $n_2=M_y$
gives the original square lattice, while the sets $n_1=M_x$ and $n_2=1$ or
$n_1=1$ and $n_2=M_y$ lead to the minimally allowed misfit angle for the
given size of the simulation cell.
Let us draw two lines (see Fig.~\ref{C16}), the first line starts at the
point O$_2 = (0, L_y)$ and has the length $n_1 a_x$, while the second line
starts at the point O$_3 = (L_x, L_y)$ and has the length $n_2 a_y$.
These lines intersect at point O$_4$ forming an angle $\pi/2 +
\varepsilon$.
The rotated substrate atoms are placed along these lines, and then periodic
shifts by multiples of $a_x$ and $a_y$ in the directions defined by these
two lines will generate the rotated oblique lattice.

\begin{figure}
\centerline{
(a)\includegraphics[width=4.cm,clip=]{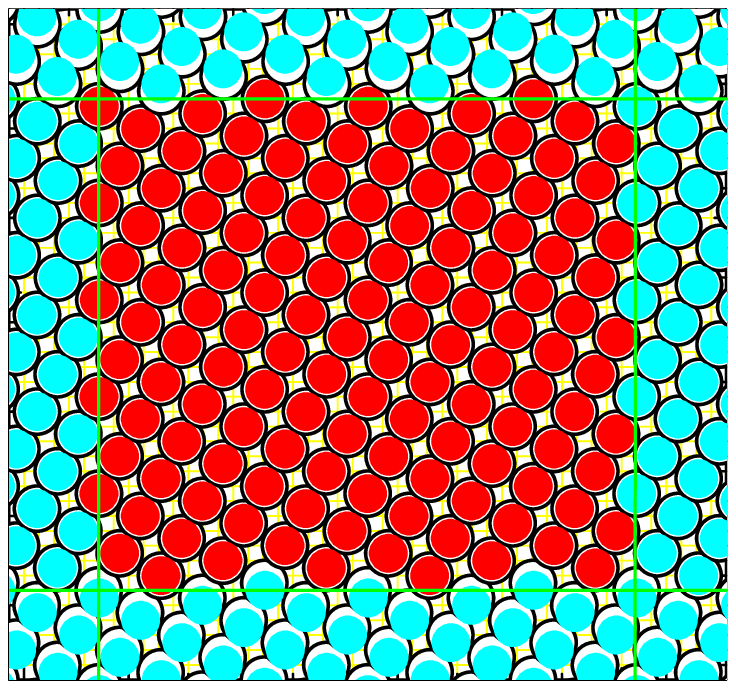}
\hfill
(b)\includegraphics[width=4.cm,clip=]{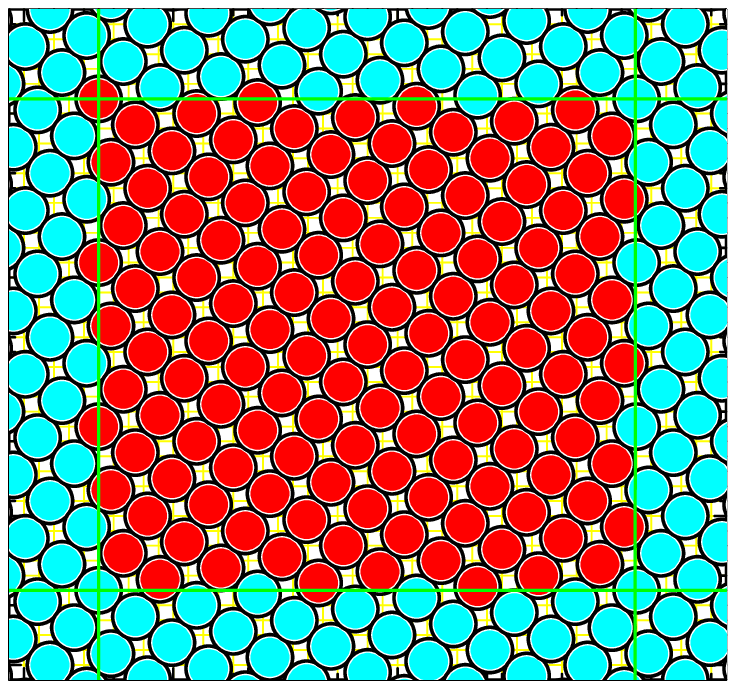}
}
\caption{\label{C17} (Color online)
Two typical examples of the rotated substrate lattice, with $M_x=12$ and
$M_y=11$:
(a) for $\phi=28.75^{\circ}$
($n_1 =10$, $n_2 =6$, $\varepsilon =0.58^{\circ}$, $a_x =3.15$, $a_y =2.86$,
$\Delta N_{\rm sub} =-2$, $\delta_1 =0.0294$), and
(b) for $\phi=34.35^{\circ}$
($n_1 =10$, $n_2 =7$, $\varepsilon =-2.3^{\circ}$, $a_x =3.01$, $a_y =2.99$,
$\Delta N_{\rm sub} =0$, $\delta_1 =10^{-9}$).
The atoms are shown by circles: those within the simulation cell are
red/dark, while their periodic ``images'' are cyan/clear.
}
\end{figure}

The oblique lattice constants $a_x$ and $a_y$ are determined by two
constrains.
First, we must preserve the area of the elementary cell:
\begin{equation}\label{axay}
a_x a_y \cos \varepsilon = a_s^2
\,.
\end{equation}
Second, from the triangle O$_2$O$_3$O$_4$ we have
\begin{equation}\label{Lx2}
L_x^2 = \left( n_1 a_x \right)^2 + \left( n_2 a_y \right)^2 +
2 n_1 \, n_2 \, a_s^2 \tan \varepsilon
\,.
\end{equation}
From Eqs.\ (\ref{axay}) and (\ref{Lx2}) we obtain
\begin{equation}\label{axfin}
\left( a_x^2 \right)_{1,2} =
\left( A \cos^2 \varepsilon \pm D \right) /
\left( 2 n_1^2 \cos^2 \varepsilon \right),
\end{equation}
where
$D = \left(
A^2 \cos^2 \varepsilon - 4 n_1^2 n_2^2 a_s^4 \right)^{1/2} \cos \varepsilon$
and
$A = L_x^2 -2 n_1 n_2 a_s^2 \tan \varepsilon$.
The signs $\pm$ in Eq.~(\ref{axfin}) yield two possible solutions which we
call as ``left'' and ``right''; typically we use the ``right'' variant when
$n_1 > M_x/2$ and the ``left'' one for $n_1 < M_x/2$.
Finally, the actual misfit angle $\phi$ is given by
$\phi = \varphi + \varepsilon /2$, where
\begin{equation}\label{phi}
\sin \varphi = n_2 a_y \cos \varepsilon / L_x.
\end{equation}

The construction described above guarantees the perfect PBC in
the $x$ direction, but not in the $y$ one.
Therefore, a next step in construction is to characterize this distortion.
Considering the triangle O$_1$O$_3$O$_5$ in Fig.~\ref{C16} (the line
O$_1$O$_5$ is parallel to O$_2$O$_4$), the lengths of its short sides are
$R_x = R_B \sin (\varphi + \varepsilon)$ and $R_y = R_B \cos \varphi$, where $R_B
= L_y / \cos \varepsilon$.
The distortion of PBC is determined by two parameters
$\delta_x = \left[ R_x/a_x - {\rm int} (R_x/a_x) \right]^2$ and
$\delta_y = \left[ R_y/a_y - {\rm int} (R_y/a_y) \right]^2$.
In the ideal case, it should be $\delta_x = \delta_y = 0$.

A ``quality'' of the rotated lattice can be characterized by two parameters:
the first parameter
\begin{equation}
\delta_1 = \delta_x^2 + \delta_y^2
\label{delta1}
\end{equation}
describes the distortion of periodic boundary conditions, while the second
parameter
\begin{equation}
\delta_2 = (a_x/a_y-1)^2
\label{delta2}
\end{equation}
indicates how close $a_x$ and $a_y$ are to the original square-lattice
constant $a_s$.
Perfect rotations are realized for $\delta_1=0$ and $\delta_2=0$
simultaneously at $\varepsilon =0$.
Thus, for given integers $n_1$ and $n_2$, we plot $\delta_1$ and $\delta_2$
as functions of $\varepsilon$, to choose an appropriate minimum of
$\delta_1 (\varepsilon)$ close to the point $\varepsilon =0$, and to check
that $\delta_2$ is not too large at that point.

Note that in the non-ideal case, some atoms within the
$L_x \times L_y$ simulation cell may be missing or, for other sets of
parameters, some atoms may overlap with their ``image'' atoms generated by
PBC.
To overcome this problem, we shifted slightly the bottom boundary of the
selected area.
As a result, in the rotated lattice the number of substrate atoms may
differ from the original one by a value $\Delta N_{\rm sub}$.

Finally, because different sets of parameters may provide approximately the
same misfit angle, we can choose the best set, the one which minimizes
$\delta_{1,2}$ and $\varepsilon$.
Two typical examples of the construction described above are shown in
Fig.~\ref{C17}.





\begin{thebibliography}{99}
\bibitem{P0} B.N.J.\ Persson, {\em ``Sliding Friction: Physical Principles and Applications''}
  (Springer-Verlag, Berlin, 1998).
\bibitem{BN2006} O.M.\ Braun and A.G.\ Naumovets, Surf.\ Sci.\ Reports {\bf 60}, 79 (2006).
\bibitem{TR1990} P.A. Thompson and M.O. Robbins, \pra {\bf 41}, 6830 (1990).
\bibitem{TR1990a} P.A. Thompson and M.O. Robbins, Science {\bf 250}, 792 (1990);
                  M.O. Robbins and P.A. Thompson, Science {\bf 253}, 916 (1991).
\bibitem{HS1990} M.\ Hirano and K.\ Shinjo, \prb {\bf 41}, 11837 (1990).
\bibitem{MR2000} M.H.\ M\"user and M.O.\ Robbins, \prb {\bf 61}, 2335 (2000).
\bibitem{HMR1999} G.\ He, M.H.\ M\"user, and M.O.\ Robbins, Science {\bf 284}, 1650 (1999).
\bibitem{MWR2000b} M.H.\ M\"user, L.\ Wenning, and M.O.\ Robbins, \prl {\bf 86}, 1295 (2001).
\bibitem{L2002} F.\ Lancon, Europhys.\ Lett.\ {\bf 57}, 74 (2002).
\bibitem{HS1993} M.\ Hirano and K.\ Shinjo, Wear {\bf 168}, 121 (1993).
\bibitem{RS1996} M.O.\ Robbins and E.D.\ Smith, Langmuir {\bf 12}, 4543 (1996).
\bibitem{Verhoeven04} G.S.\ Verhoeven, M.\ Dienwiebel, and J.W.M.\ Frenken,
  \prb \textbf{70}, 165418 (2004).
\bibitem{Bonelli09} F.\ Bonelli, N.\ Manini, E.\ Cadelano, and L.\ Colombo,
  Eur.\ Phys.\ J.\ B {\bf 70}, 449 (2009).
\bibitem{FasolinoInTrieste} A.S.\ de Wijn, C.\ Fusco, and A.\ Fasolino,
  \pre {\bf 81}, 046105 (2010).
\bibitem{GT1997} T.\ Gyalog and H.\ Thomas, Europhys.\ Lett.\ {\bf 37}, 195 (1997).
\bibitem{QCCG2002} Y.\ Qi, Y.-T.\ Cheng, T.\ Cagin, and W.A.\ Goddard III, \prb {\bf 66}, 085420 (2002).
\bibitem{HSKM1997} M.\ Hirano, K.\ Shinjo, R.\ Kaneko, and Y.\ Murata, \prl {\bf 78}, 1448 (1997).
\bibitem{DVPFHZ2004} M.\ Dienwiebel, G.S.\ Verhoeven, N.\ Pradeep, J.W.M.\ Frenken,
  J.A.\ Heimberg, and H.W.\ Zandbergen, \prl {\bf 92}, 126101 (2004).
\bibitem{HR2001s} G.\ He and M.O.\ Robbins,   \prb {\bf 64}, 035413 (2001).
\bibitem{HR2001k} G.\ He and M.O.\ Robbins, Tribology Letters {\bf 10}, 7 (2001).
\bibitem{BP2001} O.M.\ Braun and M.\ Peyrard, \pre {\bf 63}, 046110 (2001).
\bibitem{BF2002} O.M.\ Braun and R.\ Ferrando, \pre {\bf 65}, 061107 (2002).
\bibitem{BP2003} O.M.\ Braun and M.\ Peyrard, \pre {\bf 68}, 011506 (2003).
\bibitem{BraunBook} O.M.\ Braun and Yu.S.\ Kivshar,
     {\em ``The Frenkel-Kontorova Model: Concepts, Methods, and Applications''}
     (Springer-Verlag, Berlin, 2004).
\bibitem{Vanossi06} A.\ Vanossi, N.\ Manini, G.\ Divitini, G.E.\ Santoro, and E.\ Tosatti,
  \prl {\bf 97}, 056101 (2006).
\bibitem{Vanossi07PRL} A.\ Vanossi, N.\ Manini, F.\ Caruso, G.E.\ Santoro, and E.\ Tosatti,
  \prl {\bf 99}, 206101 (2007).

\bibitem{Gnecco00}
E.\ Gnecco, R.\ Bennewitz, T.\ Gyalog, Ch.\ Loppacher, M.\ Bammerlin,
E.\ Meyer, and H.-J.\ G\"untherodt, \prl {\bf 84}, 1172 (2000).

\bibitem{Sang01}
Y.\ Sang, M. Dub\'e, and M.\ Grant, \prl {\bf 87}, 174301 (2001).

\bibitem{Gnecco03}
E.\ Gnecco, R.\ Bennewitz, A.\ Socoliuc, and E.\ Meyer,
Wear {\bf 254}, 859 (2003).

\bibitem{Sills03}
S.\ Sills and R.\ M.\ Overney, \prl {\bf 91}, 095501 (2003).

\bibitem{Riedo03}
E.\ Riedo, E.\ Gnecco, R.\ Bennewitz, E.\ Meyer, and H.\ Brune,
\prl {\bf 91}, 084502 (2003).

\bibitem{Jinesh08} K.B.\ Jinesh, S.Yu.\ Krylov, H.\ Valk, M.\ Dienwiebel,
  and J.W.M.\ Frenken, \prb {\bf 78}, 155440 (2008).

\bibitem{Krylov08}
S. Yu. Krylov, and J. W. M. Frenken,
J.\ Phys.: Condens.\ Matter {\bf 20}, 354003 (2008).


\bibitem{BP2008} O.M.\ Braun and M.\ Peyrard, \prl {\bf 100}, 125501 (2008).


\bibitem{friction_rigid} N.\ Manini and O.M.\ Braun, following paper.

\bibitem{FDFKU2008} A.E.\ Filippov, M.\ Dienwiebel, J.W.M.\ Frenken,
  J.\ Klafter, and M.\ Urbakh, \prl {\bf 100}, 046102 (2008).


\end{thebibliography}
\end{document}